%% file: portmap.tex
\documentclass[sigplan,screen,letterpaper]{acmart}

\setcopyright{acmcopyright}
\acmPrice{15.00}
\acmDOI{10.1145/3385412.3385995}
\acmYear{2020}
\copyrightyear{2020}
\acmSubmissionID{pldi20main-p235-p}
\acmISBN{978-1-4503-7613-6/20/06}
\acmConference[PLDI '20]{Proceedings of the 41st ACM SIGPLAN International Conference on Programming Language Design and Implementation}{June 15--20, 2020}{London, UK}
\acmBooktitle{Proceedings of the 41st ACM SIGPLAN International Conference on Programming Language Design and Implementation (PLDI '20), June 15--20, 2020, London, UK}

\usepackage[normalem]{ulem} 

\usepackage[english]{babel}
\addto\extrasenglish{%
}
\usepackage[utf8]{inputenc}
\usepackage{csquotes}     
\usepackage{xspace}       
\usepackage{nag}          
\usepackage{tikz}         
\usepackage{siunitx}
\usetikzlibrary{calc}
\usetikzlibrary{positioning}
\usetikzlibrary{patterns}
\usepackage{pgf}          
\usepackage{algorithm2e}  
\usepackage{natbib}       
\usepackage{amsmath}      
\usepackage{amsthm}       
\usepackage{subcaption} 
\usepackage{booktabs}   

\SetAlgorithmName{Algorithm}{Algorithm}{Algorithm}
\SetAlgoCaptionSeparator{.}

\SetAlCapSkip{0.6em}

\input{tex/definitions}

\newcommand{\tool}{PMEvo\xspace}

\begin{document}

\title{PMEvo: Portable Inference of \PortMappings for Out-of-Order Processors by Evolutionary Optimization}


\author{Fabian Ritter}
\affiliation{
  \institution{Saarland University\\Saarland Informatics Campus}            
  \country{Germany}                    
}
\email{fabian.ritter@cs.uni-saarland.de}          

\author{Sebastian Hack}
\orcid{nnnn-nnnn-nnnn-nnnn}             
\affiliation{
  \institution{Saarland University\\Saarland Informatics Campus}            
  \country{Germany}                    
}
\email{hack@cs.uni-saarland.de}          

\begin{abstract}
  \noindent
  \input{tex/abstract}

\end{abstract}



\maketitle

\input{tex/intro}

\input{tex/background}

\input{tex/procmodel}

\input{tex/evoalgo}

\input{tex/eval}

\input{tex/relwork}

\input{tex/conclusion}


\appendix
\input{tex/correctness}

\bibliography{references.bib}

\end{document}

%% file: tex/definitions.tex
\def\defset#1{\expandafter\def\csname#1\endcsname{\ensuremath{\mathbf{#1}}\xspace}}
\def\deffun#1{\expandafter\def\csname#1\endcsname{\ensuremath{\mathit{#1}}\xspace}}

\makeatletter
\def\moverlay{\mathpalette\mov@rlay}
\def\mov@rlay#1#2{\leavevmode\vtop{%
   \baselineskip\z@skip \lineskiplimit-\maxdimen
   \ialign{\hfil$\m@th#1##$\hfil\cr#2\crcr}}}
\newcommand{\charfusion}[3][\mathord]{
    #1{\ifx#1\mathop\vphantom{#2}\fi
        \mathpalette\mov@rlay{#2\cr#3}
      }
    \ifx#1\mathop\expandafter\displaylimits\fi}
\makeatother

\newcommand{\cupdot}{\charfusion[\mathbin]{\cup}{\cdot}}

\newcommand{\disjointunion}{\cupdot}

\newcommand{\reals}{\ensuremath{\mathbb{R}}\xspace}
\newcommand{\nat}{\ensuremath{\mathbb{N}}\xspace}

\theoremstyle{definition}
\newtheorem{definition}{Definition}

\newtheorem{example}{Example}

\newcommand{\fspace}{\quad\quad}

\def\eg{e.g.\@\xspace}
\def\ie{i.e.\@\xspace}

\newcommand{\insn}[1]{\ensuremath{\mathit{#1}}\xspace}

\newcommand{\insnf}[1]{#1}

\newcommand{\Insns}{\ensuremath{\mathbf{I}}}
\newcommand{\Ports}{\ensuremath{\mathbf{P}}}
\newcommand{\Uops}{\ensuremath{\mathbf{U}}}

\newcommand{\tpmeas}[1]{\ensuremath{t^*(#1)}\xspace}
\newcommand{\tpsim}[2]{\ensuremath{t^*_{#1}(#2)}\xspace}

\newcommand{\uop}{$\mu$op\@\xspace}
\newcommand{\uops}{$\mu$ops\@\xspace}

\newcommand{\PortMappings}{Port Mappings\@\xspace}

\newcommand{\portmapping}{port mapping\@\xspace}
\newcommand{\portmappings}{port mappings\@\xspace}
\newcommand{\portmappingbound}{port-mapping-bound\@\xspace}

\newcommand{\bnalgo}{bottleneck simulation algorithm\@\xspace}

\newcommand{\tikzinsn}[2]{ \node[anchor=south] (i#1) at (#1,0){#2}; }
\newcommand{\tikzport}[2]{ \node[anchor=south] (p#1) at ($(#1,-1.4) + (\tikzportoff, 0)$){#2}; }
\newcommand{\tikzmaps}[2]{ \draw[] (i#1) to (p#2); }
\newcommand{\tikzportoff}{0}

\newcommand{\Utikzinsn}[3]{ \node[anchor=south] (i#1) at (#2,0){#3}; }
\newcommand{\Utikzuop}[3]{ \node[anchor=south] (u#1) at (#2,-1.2){#3}; }
\newcommand{\Utikzport}[3]{ \node[anchor=south] (p#1) at (#2,-2.0){#3}; }

\newcommand{\Utikzmapsitou}[3]{ \draw[] (i#1) -- node[left] {#3} (u#2); }
\newcommand{\Utikzmapsutop}[2]{ \draw[] (u#1) to (p#2); }

\newcommand{\myCyan}{cyan!80!black}
\newcommand{\myGreen}{green!80!black}
\newcommand{\myOrange}{orange!80!white}

\newcommand{\myvscale}{2}

\newcommand{\drawbucket}[4]{
  \draw[thick] ($(#1) + (0,\myvscale * #2)$) -- ($(#1)$) -- ($(#1) + (#3,0)$) -- ($(#1) + (#3,\myvscale * #2)$);
  \node[] at ($(#1) + (0.5 * #3,-0.4)$) {#4};
}

\newcommand{\drawinsn}[5]{
  \draw[#5, thick] let \p{A}=(#1) in
  ($(\x{A}, \myvscale * \y{A})$) rectangle ($(\x{A}, \myvscale * \y{A}) + (#3, \myvscale * #2)$) node[pos=.5] {#4};
}

\newcommand{\drawscale}[2]{
  \draw[thick] (#1) -- ($(#1) + (0, \myvscale * #2)$);
  \foreach \y in {0,...,#2}
    \draw[thick] ($(#1) + (2pt,\myvscale * \y)$) -- ($(#1) + (-2pt,\myvscale * \y)$) node[anchor=east] {$\y$};
}

\let\pgfimageWithoutPath\pgfimage
\renewcommand{\pgfimage}[2][]{\pgfimageWithoutPath[#1]{pics/#2}}

%% file: tex/abstract.tex
Achieving peak performance in a computer system requires optimizations in every layer of the system, be it hardware or software.
A detailed understanding of the underlying hardware, and especially the processor, is crucial to optimize software.
One key criterion for the performance of a processor is its ability to exploit instruction-level parallelism.
This ability is determined by the \portmapping of the processor, which describes the execution units of the processor for each instruction.

Processor manufacturers usually do not share the \portmappings of their microarchitectures.
While approaches to automatically infer \portmappings from experiments exist, they are based on processor-specific hardware performance counters that are not available on every platform.

We present \tool, a framework to automatically infer \portmappings solely based on the measurement of the execution time of short instruction sequences.
\tool uses an evolutionary algorithm that evaluates the fitness of candidate mappings with an analytical throughput model formulated as a linear program.
Our prototype implementation infers a \portmapping for Intel's Skylake architecture that predicts measured instruction throughput with an accuracy that is competitive to existing work.
Furthermore, it finds \portmappings for AMD's Zen+ architecture and the ARM Cortex-A72 architecture, which are out of scope of existing techniques.

%% file: tex/intro.tex
\section{Introduction}
\label{sec:intro}

Accurately estimating the time required to execute a given program has become increasingly complex.
While advances in hardware design enable faster execution times, they make it difficult to optimize programs such that they utilize the available resources to the best possible extent.
A particular cause of unforeseen performance characteristics is the exploitation of instruction level parallelism via out-of-order execution~\citep{tomasulo67}.
This technique enables the processor to dynamically re-order the instructions of a sequential program and execute them in parallel on a set of execution ports.
Therefore, optimizing a program to achieve peak performance on a processor requires knowledge of the ports that can be used by each instruction.
However, the instruction-to-port mapping, or \portmapping, is usually only known to hardware manufacturers and may vary with each new hardware generation.

While approaches towards understanding the performance characteristics of processors without full insight into their internals exist, they suffer from shortcomings:
Some approaches require significant manual effort \citep{llvmmca, agner18} or are restricted to validating existing \portmappings \citep{laukemann18}.
Others are closely tied to microarchitecture-specific performance counters \citep{abel19, agner18, exegesis} that prevent their applicability to a wide range of practically-relevant processors.
Another line of research~\citep{mendis19} uses machine learning to train a neural network that estimates instruction throughput.
This approach is portable among microarchitectures, but the resulting black box model is hard to use for identifying concrete performance bottlenecks.

This paper proposes a solution that comes without any of these drawbacks:
Experiments are automatically generated from a description of the available instructions.
Performing the experiments requires only measuring the time taken for executing an instruction sequence.
The result is a concise and interpretable \portmapping model that existing tools can use to identify bottlenecks and to guide optimization decisions.

We achieve this with \tool, a framework that uses an evolutionary algorithm to find a \portmapping that excels in explaining measured throughputs for automatically generated instruction sequences.
These instruction sequences are designed to reveal conflicting resource requirements for pairs of instructions while exhibiting as few data dependencies as possible.
For these instruction sequences, the throughput is only limited by constrained ports and therefore carries information about the \portmapping.

A key component of the evolutionary algorithm is a novel \bnalgo to evaluate the fitness of candidate \portmappings.
This algorithm efficiently computes the solution of a linear program that models an optimal instruction scheduler for a given \portmapping.
Our novel \bnalgo outperforms solving the corresponding linear program for realistic \portmappings by two orders of magnitude.

We evaluate \tool by the throughput prediction accuracy of its inferred \portmappings on \portmappingbound basic blocks for microarchitectures by Intel, AMD and ARM.
\tool's prediction accuracy for the Intel Skylake architecture is close to existing approaches like IACA \citep{iaca} and uops.info \citep{abel19} that rely on stronger knowledge about the microarchitecture.
For AMD and ARM, \tool outclasses the state-of-the-art \portmapping model of llvm-mca~\citep{llvmmca}.

In summary, we make the following contributions:
\begin{itemize}
  \item
    An evolutionary algorithm that infers \portmappings from specifically designed experiments with measured throughputs without relying on microarchitecture-specific features.

  \item
    A \bnalgo that allows to efficiently evaluate the fitness of \portmappings for a given set of experiments.

  \item
    A prototype implementation that finds a \portmapping that is competitive to related work for Intel's Skylake architecture and that is the first one to automatically find \portmappings for the AMD Zen+ and the ARM Cortex-A72 microarchitectures.
\end{itemize}


%% file: tex/background.tex
\section{Background: Processor Design}
\label{sec:background}

Modern processors apply out-of-order execution~\citep{tomasulo67}.\footnote{A contemporary introduction to the topic can be found \eg in Chapter 3 of the textbook by \citet{hennessy11}.}
This concept is based on the observation that instructions can be executed in any order as long as the results are the same as if they were executed in program order.
Therefore, a processor may execute instructions in parallel and reorder them to any extent that preserves the read-after-write dependencies between the operations and the externally-visible effects.

Out-of-order execution is often combined with a scheme to decompose instructions into simpler microarchitecture-specific operations.
These so-called micro-ops or \uops are then subject to reordering.

\begin{figure}
    \centering
    \scalebox{.9}{
      \begin{tikzpicture}
        \draw[rounded corners] (0.2,0) rectangle (3,-0.6);
        \node[] at (1.6,-0.3) (icache) {L1 ICache};

        \draw[-latex] (1.6, -0.6) -- (1.6, -1.0);

        \draw[rounded corners] (-3,-1.0) rectangle (-0.2,-1.6);
        \node[anchor=north] at (-1.6,-1.05) (uopcache) {\uop Cache};

        \draw[-latex] (-1.6, -1.6) -- (-1.6, -2.0);

        \draw[rounded corners] (0.2,-1.0) rectangle (3,-1.6);
        \node[anchor=north] at (1.6,-1.05) (decode) {Decoder};

        \draw[-latex] (1.6, -1.6) -- (1.6, -2.0);
        \draw[-latex] (0.2, -1.3) -- (-0.2, -1.3);

        \draw[rounded corners] (-3,-2.0) rectangle (3,-2.6);
        \node[] at (0,-2.3) (reg) {Register Management};

        \draw[-latex] (0.0, -2.6) -- (0.0, -3.0);

        \draw[rounded corners] (-3,-3.0) rectangle (3,-3.6);
        \node[] at (0,-3.3) (sched) {Scheduler};

        \draw[-latex] (-2.325, -3.6) -- (-2.325, -4.0);
        \draw[-latex] (-0.775, -3.6) -- (-0.775, -4.0);
        \draw[-latex] (0.775, -3.6) -- (0.775, -4.0);
        \draw[-latex] (2.325, -3.6) -- (2.325, -4.0);

        \draw[] (-3,-4.0) rectangle (-1.65,-4.6);
        \node[] at (-2.325,-4.3) (p0) {Port 0};

        \draw[] (-3,-4.6) rectangle (-1.65,-6.0);

        \node[anchor=north, align=center] at (-2.325,-4.6) {\small Int ALU\\\small Vec ALU\\\small DIV};

        \draw[] (-1.45,-4.0) rectangle (-0.1,-4.6);
        \node[] at (-0.775,-4.3) (p1) {Port 1};

        \draw[] (-1.45,-4.6) rectangle (-0.1,-6.0);

        \node[anchor=north, align=center] at (-0.775,-4.6) {\small Int ALU\\\small Vec ALU};

        \draw[] (0.1,-4.0) rectangle (1.45,-4.6);
        \node[] at (0.775,-4.3) (p2) {Port 2};

        \draw[] (1.45,-4.6) rectangle (0.1,-6.0);

        \node[anchor=north, align=center] at (0.775,-4.6) {\small LD/ST};

        \draw[] (1.65,-4.0) rectangle (3,-4.6);
        \node[] at (2.325,-4.3) (p3) {Port 3};

        \draw[] (3,-4.6) rectangle (1.65,-6.0);

        \node[anchor=north, align=center] at (2.325,-4.6) {\small ST};

        \draw[latex-latex] (0.775, -6.0) -- (0.775, -6.4);
        \draw[-latex] (2.325, -6.0) -- (2.325, -6.4);

        \draw[rounded corners] (0.2,-6.4) rectangle (3,-7.0);
        \node[] at (1.6,-6.7) (dcache) {L1 DCache};
      \end{tikzpicture}
    }
    \Description{
      Simplified overview of a modern processor design.
      The relevant features are described in the section text.
    }
    \caption{
      Simplified overview of a modern processor design (based on Figure 2-3 in the Intel Software Optimization Manual~\citep{intel19})
    }
    \label{fig:proc_schematic}
  \end{figure}
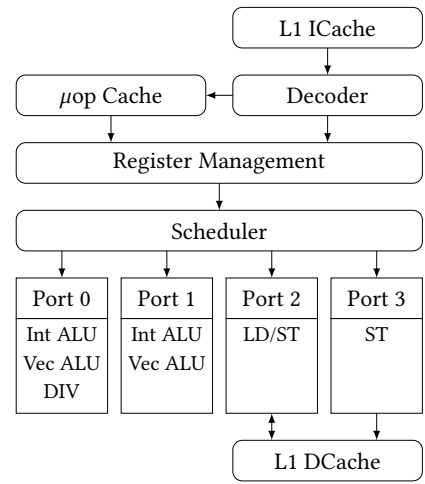

\autoref{fig:proc_schematic} shows the relevant parts of a microarchitecture that employs out-of-order execution and \uop decomposition.
Instructions are fetched and decoded from the instruction cache in program order.
The decoder produces \uops, which are cached for future re-use.
The register management engine resolves false (write-after-read or write-after-write) dependencies by mapping the operand registers of each operation to a larger number of physical registers.
A scheduler decides based on operand dependencies and resource availability when and where to execute the \uops.
The execution units (\eg arithmetical units and load/store units), which execute the \uops, are grouped behind ports.
Often, execution units are pipelined, allowing the ports to start processing a new instruction in every cycle.
Several instances of the same kind of execution unit can exist at different ports.

A key factor to the running time of a given piece of code on a processor is therefore the \portmapping.
It specifies how instructions are decomposed into \uops and which \uops can be executed on which ports.\footnote{This is called \emph{port usage} in the work by \citet{abel19}.}

Some microarchitectures, particularly those designed by Intel, provide fine-grained hardware performance counters that count the number of executed \uops per port.
While these greatly help at inferring \portmappings, relying on them excludes all microarchitectures that do not provide similar performance counters.
Therefore, we base our approach on the more portable observation of throughputs as defined in the following.\footnote{This definition is equivalent to the one by \citet{mendis19}, an extension of the instruction-wise throughput definition used by \citet{agner18} and \citet{abel19}.}

\begin{definition}
  \label{def:throughput}
  The \emph{throughput}~$\tpmeas{e}$ of an instruction sequence (or experiment)~$e$ on a given processor is the average number of processor cycles required to execute~$e$ in a steady state.
\end{definition}
The execution of an experiment in an infinite loop is considered to have reached a steady state when the average number of required cycles per iteration stays constant for the remaining execution.

%% file: tex/procmodel.tex
\section{Analytical Throughput Model}
\label{sec:procmodel}

Since our goal is to infer a \portmapping from throughput measurements, we need to understand the connection between the \portmapping of the processor and the throughput that is achieved for an experiment.
The precise inner workings of processors are well-kept secrets of the manufacturers, therefore we postulate a model of how processors execute instructions with respect to a \portmapping.
In this section, we present a model that is supported by information provided by hardware manufacturers \citep{amd17,arm15,intel19,iaca} and related research~\citep{abel19,agner18}.
The linear programs we use in this section to define the throughput for a given \portmapping are extensions of work presented by \citet{abel19}.

\subsection{Out-of-Order Throughput Model}
\label{ssec:two_level_mapping}

We start by defining a simple out-of-order execution model that does not consider the decomposition of instructions into \uops.
We refer to it as the \emph{two-level model} (mapping only instructions to ports).
\autoref{ssec:uop_decomp} extends this model to the \emph{three-level model}, which additionally supports decomposing instructions into \uops.

Our throughput model is based on the notion of a \portmapping, as defined in the following.

\begin{definition}
  A \emph{\portmapping in the two-level model} is a bipartite graph $(\Insns \disjointunion \Ports, M)$ with the nodes split disjointly into a set~$\Insns$ of instructions and a set~$\Ports$ of ports and edges $M\subseteq \Insns\times\Ports$ between these.
\end{definition}
An edge between instruction~$i$ and port~$k$ indicates that instruction~$i$ can be executed on port~$k$.

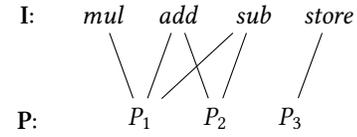
\begin{figure}[ht]
  \centering
  \begin{tikzpicture}
    \renewcommand{\tikzportoff}{.5}
    \node[anchor=south] (icap) at (0, 0) {\Insns:};
    \node[anchor=south] (pcap) at (0, -1.4) {\Ports:};
    \tikzinsn{1}{\insn{mul}}
    \tikzinsn{2}{\insn{add}}
    \tikzinsn{3}{\insn{sub}}
    \tikzinsn{4}{\insn{store}}

    \tikzport{1}{$P_1$}
    \tikzport{2}{$P_2$}
    \tikzport{3}{$P_3$}

    \tikzmaps{2}{1}
    \tikzmaps{2}{2}
    \tikzmaps{3}{1}
    \tikzmaps{3}{2}
    \tikzmaps{1}{1}
    \tikzmaps{4}{3}
  \end{tikzpicture}
  \Description{
    Example of a \portmapping in the two-level model.
    It is depicted as a bipartitie graph mapping the instructions as described in the section text.
  }
  \caption{
    Example of a \portmapping in the two-level model
  }
  \label{fig:ex_two_level}
\end{figure}

Consider the \portmapping shown in \autoref{fig:ex_two_level}.
There are four instructions \insn{mul}, \insn{add}, \insn{sub}, and \insn{store} that are mapped to the three ports $P_1$, $P_2$, and $P_3$.
The two instructions \insn{add} and \insn{sub} can both be executed on the same two ports $P_1$ and $P_2$, \insn{mul} can use only one of them, $P_1$, and \insn{store} has to be executed on a separate port $P_3$.

In this model, an experiment is represented as a multiset of instructions, \ie a function $e:\Insns\rightarrow\nat$ that maps instructions to their number of occurrences.
We abstract from the order of the instructions since we only use experiments that can be reordered freely by the scheduler.
The throughput of an experiment with a given \portmapping is characterized by the following definiton.

\begin{definition}
  \label{def:tp2l}
  Given a \portmapping $m := (\Insns \disjointunion \Ports, M)$ in the two-level setting, the \emph{throughput}~$\tpsim{m}{e}$ under $m$ for an experiment~$e:\Insns\rightarrow\nat$ is the objective value of an optimal solution to the following linear program:
\begin{alignat*}{3}
  &\text{minimize}& \quad & t \\
  &\text{subject to}& \quad &\displaystyle\sum_{k\in\Ports} x_{ik} = \insnf{e}(i)  && \fspace\text{for all $i\in\Insns$}\tag{A}\label{jsA}\\
  &&&  \displaystyle\sum_{i\in\Insns} x_{ik} \le t      && \fspace\text{for all $k\in\Ports$}\tag{B}\label{jsB}\\
  &&&  x_{ik} \ge 0                          && \fspace\text{for all $(i, k) \in M$}\tag{C}\label{jsC}\\
  &&&  x_{ik} = 0                            && \fspace\text{for all $(i, k) \not\in M$}\tag{D}\label{jsD}
\end{alignat*}%
\end{definition}
The intuition for this linear program is that each instruction~$i$ in the experiment has the mass $\insnf{e}(i)$.
This mass is distributed among the ports that can execute~$i$, as required by constraint~(\ref{jsA}).
The $x_{ik}$ are real-valued variables that represent the share of the mass $\insnf{e}(i)$ that is executed on port~$k$ in the experiment.
Constraint~(\ref{jsB}) establishes the objective $t$ as an upper bound of the sums of mass shares on each port.
The constraints~(\ref{jsC}) and~(\ref{jsD}) guarantee that the mass of an instruction $i$ is distributed to ports that can execute~$i$.
The throughput is the maximal mass associated to any port if all mass is distributed as evenly as possible.

\begin{example}
  \label{ex:tp}
  \autoref{fig:ex_throughput} displays a graphical interpretation of an optimal solution of the linear program for the experiment
  \[
    e := \{\insn{add}\mapsto 2, \insn{mul}\mapsto 1, \insn{store}\mapsto 1\}
  \]
  under the mapping given in \autoref{fig:ex_two_level}.

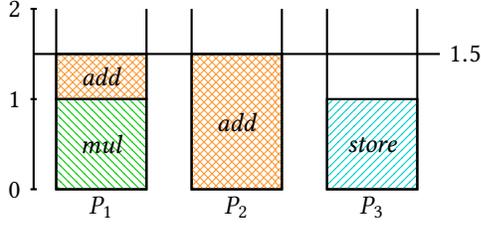
\begin{figure}[ht]
  \centering
  \begin{tikzpicture}[scale=.6]
    \drawbucket{0,0}{2}{2}{$P_1$}
    \drawbucket{3,0}{2}{2}{$P_2$}
    \drawbucket{6,0}{2}{2}{$P_3$}

    \drawinsn{0,0}{1}{2}{\insn{mul}}{pattern color=\myGreen,pattern=north west lines}
    \drawinsn{0,1}{0.5}{2}{\insn{add}}{pattern color=\myOrange,pattern=crosshatch}
    \drawinsn{3,0}{1.5}{2}{\insn{add}}{pattern color=\myOrange,pattern=crosshatch}
    \drawinsn{6,0}{1}{2}{\insn{store}}{pattern color=\myCyan,pattern=north east lines}

    \drawscale{-.5,0}{2}

    \draw[thick] ($(-0.5, \myvscale * 1.5)$) -- ($(8.5, \myvscale * 1.5)$);
    \node[anchor=west] at ($(8.5, \myvscale* 1.5)$) {$1.5$};
  \end{tikzpicture}
  \Description{
    Visualization of an example port allocation.
    One unit of mul instructions is placed on port 1, together with a half unit of add instructions.
    The remaining $1.5$ units of add instructions is on port 2.
    Port 3 only contains 1 unit of store instructions.
  }
  \caption{Visualization of an example port allocation}
  \label{fig:ex_throughput}
\end{figure}

The mass allocated to each port $P_k$ is drawn in the corresponding bucket.
The throughput of $1.5$ cycles is the mass of the most occupied ports~$P_1$ and~$P_2$\@.
Note that the mass of the two \insn{add} instructions is split unevenly among two ports.
While these non-integer instruction portions might seem counter-intuitive, they assort with the definition of throughput as the average number of cycles to execute an experiment.

Such a throughput can be realized by executing one \insn{add} instruction on port $P_1$ in every second iteration, yielding an average of~$0.5$ \insn{add} instructions on~$P_1$ per execution of the experiment.
%
%
\end{example}

\autoref{def:tp2l} relies on several assumptions:
\begin{enumerate}
  \item The processor schedules the instructions optimally. \label{it1:opt}
  \item Operational units are fully pipelined, \ie every instruction blocks exactly one port for exactly one cycle.\label{it1:pipelined}
  \item The fetch and decode units do not impose a bottleneck for the experiment. \label{it1:frontend}
  \item There are no relevant (read-after-write) data dependencies among the instructions of the experiment. \label{it1:indep}
\end{enumerate}
The validity of assumptions~(\ref{it1:opt}) and~(\ref{it1:pipelined}) depends on the processor under test.
We found these to be usually fulfilled by modern microarchitectures.
An exception are complex instructions like divisions, which can block a port or an operational unit for multiple cycles.

We ensure the validity of assumptions~(\ref{it1:frontend}) and~(\ref{it1:indep}) by selecting our experiments appropriately:
Only sufficiently short experiments that do not hit bottlenecks in the fetch/decode stages are considered.
The operands of the experiments are furthermore chosen such that writing instructions can be retired before their output register is read for the next time.

\subsection{Micro-Operation Decomposition}
\label{ssec:uop_decomp}

To represent the decomposition of instructions into \uops, we extend the definition of a \portmapping by a layer of \uops as follows.

\begin{definition}
  A \emph{\portmapping in the three-level model} is a tripartite graph $(\Insns\disjointunion\Uops\disjointunion\Ports, N\disjointunion M)$ with labeled edges $N\subseteq\Insns\times\nat\times\Uops$ between instructions and \uops as well as unlabeled edges $M\subseteq \Uops\times\Ports$ between \uops and ports.
\end{definition}
A labeled edge~$(i, n, u)\in N$ means that there are~$n$ instances of the \uop~$u$ in the \uop decomposition of instruction~$i$.

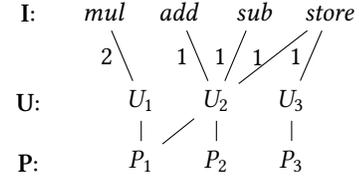
\begin{figure}[ht]
  \centering
  \begin{tikzpicture}
    \renewcommand{\tikzportoff}{.5}
    \node[anchor=south] (icap) at (-2.5, 0) {\Insns:};
    \node[anchor=south] (ucap) at (-2.5, -1.2) {\Uops:};
    \node[anchor=south] (pcap) at (-2.5, -2.0) {\Ports:};
    \Utikzinsn{1}{-1.5}{\insn{mul}}
    \Utikzinsn{2}{-0.5}{\insn{add}}
    \Utikzinsn{3}{0.5}{\insn{sub}}
    \Utikzinsn{4}{1.5}{\insn{store}}

    \Utikzuop{1}{-1}{$U_1$}
    \Utikzuop{2}{0}{$U_2$}
    \Utikzuop{3}{1}{$U_3$}

    \Utikzport{1}{-1}{$P_1$}
    \Utikzport{2}{0}{$P_2$}
    \Utikzport{3}{1}{$P_3$}

    \Utikzmapsitou{1}{1}{2}
    \Utikzmapsitou{2}{2}{1}
    \Utikzmapsitou{3}{2}{1}
    \Utikzmapsitou{4}{2}{1}
    \Utikzmapsitou{4}{3}{1}

    \Utikzmapsutop{1}{1}
    \Utikzmapsutop{2}{1}
    \Utikzmapsutop{2}{2}
    \Utikzmapsutop{3}{3}
  \end{tikzpicture}
  \Description{
    Example of a \portmapping in the two-level model.
    It is depicted as a tripartitie graph mapping the instructions as described in the section text.
  }
  \caption{
    Example of a three-level \portmapping
  }
  \label{fig:ex_three_level}
\end{figure}

An example is displayed in \autoref{fig:ex_three_level}.
Here, \insn{add} and \insn{sub} are implemented as one \uop $U_2$ that can be executed on two ports $P_1$ and $P_2$.
The \insn{mul} and \insn{store} instructions are decomposed into two \uops, the former in two of the same kind,~$U_1$, and the latter into two different ones,~$U_2$ and~$U_3$.
The \insn{store} instruction has a partial conflict with \insn{add} and \insn{sub} that cannot be represented in the two-level model.

It is important to notice the different semantics of the layers of edges:
For each instance of an instruction~$i$, \emph{all} corresponding \uops $u$ such that $(i, \cdot, u)\in N$ have to be executed whereas a \uop $u$ is executed on \emph{exactly one} of the allowed ports $k$ such that $(u, k)\in M$.
The linear program from \autoref{ssec:two_level_mapping} can be slightly modified to compute the throughput $\tpsim{m}{e}$ of an experiment~$e:\Insns\rightarrow\nat$ under the \emph{three-level} \portmapping $m := (\Insns\disjointunion\Uops\disjointunion\Ports, N\disjointunion M)$:
\begin{alignat*}{3}
  &\text{minimize}& \quad & t \\
  &\text{subject to}& \quad &\displaystyle\sum_{k\in\Ports} x_{uk} = \hspace{-0.6em}\displaystyle\sum_{(i, n, u)\in N} \hspace{-0.8em}\insnf{e}(i) \cdot n && \quad\text{for all $u\in \Uops$}\tag{A}\label{js3A}\\
  &&&  \displaystyle\sum_{u\in\Uops} x_{uk} \le t      && \quad\text{for all $k\in\Ports$}\tag{B}\label{js3B}\\
  &&&  x_{uk} \ge 0                          && \quad\text{for all $(u, k) \in M$}\tag{C}\label{js3C}\\
  &&&  x_{uk} = 0                            && \quad\text{for all $(u, k) \not\in M$}\tag{D}\label{js3D}
\end{alignat*}
All previous occurrences of instructions are replaced by occurrences of \uops except for the right-hand side of constraint~\eqref{js3A}.
The right-hand side of \eqref{js3A} ensures that a \uop $u$ that occurs $n$ times in the decomposition of instruction~$i$ is taken into account with its appropriate mass.

A valuable observation is that computing the throughput of an experiment~$e:\Insns\rightarrow\nat$ with a \portmapping $(\Insns\disjointunion\Uops\disjointunion\Ports, N\disjointunion M)$ in the three-level model can be reduced to computing throughput in the simpler two-level model:
We instead compute the throughput of the experiment
\[
  e' = \bigg\{u\mapsto \sum_{(i, n, u)\in N} \insnf{e}(i) \cdot n \bigg\}
\]
with the two-level mapping $(\Uops\disjointunion\Ports, M)$.
The multiset $e'$ contains the \uops that are needed to execute $e$ according to $N$.
These \uops are used as instructions for the two-level model.

This construction allows us to use an algorithm for the simpler two-level model to compute throughput in the three-level model.

%% file: tex/evoalgo.tex
\section{The \tool Framework}
\label{sec:evoalgo}

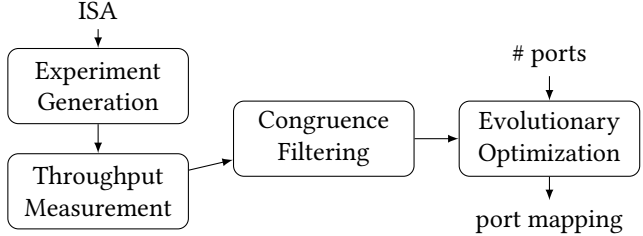
\begin{figure}[ht]
  \centering
  \begin{tikzpicture}
    \node[] (start) at (3,1.4) {ISA};
    \node[] (ports) at (9,0.80) {\# ports};
    \node[draw=black, rectangle, rounded corners, align=center, minimum width=2.4cm, minimum height=1.0cm] (gen) at (3,0.4) {Experiment\\Generation};
    \node[draw=black, rectangle, rounded corners, align=center, minimum width=2.4cm, minimum height=1.0cm] (eval) at (3,-1) {Throughput\\Measurement};
    \node[draw=black, rectangle, rounded corners, align=center, minimum width=2.4cm, minimum height=1.0cm] (preproc) at (6,-0.30) {Congruence\\Filtering};
    \node[draw=black, rectangle, rounded corners, align=center, minimum width=2.4cm, minimum height=1.0cm] (evo) at (9,-0.30) {Evolutionary\\Optimization};
    \node[] (end) at (9,-1.40) {\portmapping};
    \draw[-latex] (start) -- (gen);
    \draw[-latex] (gen) -- (eval);
    \draw[-latex] (eval) -- (preproc);
    \draw[-latex] (preproc) -- (evo);
    \draw[-latex] (evo) -- (end);
    \draw[-latex] (ports) -- (evo);
  \end{tikzpicture}
  \Description{
    \tool framework overview.
    The process starts by giving an ISA specification to the Experiment Generation phase of \tool.
    The results are given to the Throughput Measurement phase, which hands its results to the Congruence Filtering stage.
    The resulting filtered instruction forms are given to the Evolutionary Optimization phase together with the number of ports to assume for the microarchitecture.
    The result is a port mapping.
  }
  \caption{\tool framework overview}
  \label{fig:tooloverview}
\end{figure}

We propose the \tool framework to automatically infer \portmappings from throughput experiments.
An overview of this framework is given in \autoref{fig:tooloverview}.

\tool consists of four main stages, which we describe in the following subsections: Generating relevant experiments~(\ref{ssec:expsampling}), measuring the throughput of the experiments on a given processor~(\ref{ssec:expeval}), a preprocessing step that identifies congruent instructions~(\ref{ssec:preprocessing}), and evolutionary optimization~(\ref{ssec:evo}).


\subsection{Experiment Generation}
\label{ssec:expsampling}

The input of the first stage of \tool is a description of the instruction set architecture under test.
This description is a set of instruction forms, \ie instructions with typed placeholders for their operands.
The type of the placeholder specifies the operand kind (\eg memory operand, general purpose or vector register) and the width of the respective operand.
There can be multiple instruction forms for the same operation with different operand types.

\tool constructs a set of experiments from this information with the following components:
\begin{enumerate}
  \item for each instruction form $i$, an experiment $\{i\mapsto 1\}$ measuring its individual throughput $\tpmeas{i}$
  \item for each pair $(i_A, i_B)$ of instruction forms, an experiment $\{i_A\mapsto 1, i_B\mapsto 1\}$
  \item for each pair $(i_A, i_B)$ of instruction forms with $\tpmeas{i_A} >\tpmeas{i_B}$, an experiment $\{i_A\mapsto 1, i_B\mapsto n\}$ where \[n = \left\lceil \tpmeas{i_A} / \tpmeas{i_B}\right\rceil\]
\end{enumerate}

Experiments with this structure lead to different outcomes depending on the \portmapping:
If the \uops of two instruction forms $i_A$ and $i_B$ require the same resources, experiment (2) will result in a throughput that is the sum of the individual throughputs of $i_A$ and $i_B$.
In case the \uops of $i_A$ and $i_B$ are executed by disjoint execution units, the throughput of experiment (3) will be $n \cdot \tpmeas{i_B}$.
More complex partial resource conflicts will lead to measured throughputs for these experiments that are harder to interpret manually.
It is the task of the evolutionary algorithm to find a mapping that explains these throughputs.

The evolutionary algorithm is not restricted to experiments of this structure.
In theory, longer experiments that combine instances of more than two different instruction forms can unveil resource conflicts that cannot be covered by these experiments.
However, when exploring the experiment design space experimentally for existing processors, we did not observe benefits in \portmapping quality from more complex experiments.

\subsection{Throughput Measurement}
\label{ssec:expeval}

The goal of this stage is to measure the throughput of the generated experiments.
Our measurement method follows \autoref{def:throughput}:
The instruction forms of the experiment are instantiated with operands while avoiding data dependencies.
The resulting instruction sequence is executed in a loop such that the execution reaches a steady state.

\tool uses a register allocator that assigns a register from the appropriate register class to each register operand of the instruction forms.
To avoid harmful dependencies, written operands are instantiated with most recently read registers and read operands with least recently written registers.
Using as many different registers as available, this ensures that instructions with long latencies have enough time to complete before their results are read.

Memory operands are instantiated with a separate register containing a valid base pointer and one of several different constant offsets to avoid data dependencies on the memory.

Before operand allocation, we unroll several loop iterations.
This has several benefits:
It further increases the dependence distance by allowing more registers to be allocated and it avoids loop-carried dependencies.
Additionally, it reduces the influence of the loop code on our time measurements.
The range of loop body lengths that achieve an optimal throughput depends on the microarchitecture.
We found a length of 50 instructions to be in the appropriate range for all of our evaluated architectures.
With this length, the loop body will be resident in the \uop cache (if the architecture has one).
This avoids performance bottlenecks due to restrictions on the number of concurrently fetched and decoded instructions.
The loop bound is automatically chosen to ensure that the loop runs for a specific time that guarantees steady-state execution. 
This time is estimated empirically for the processor under test by comparing the measurement stability for different times.
For the evaluated platforms, we found a time of $\SI{10}{\milli\second}$ to be appropriate.

To measure the throughput, we emit the instantiated loop as inline assembly into a C program wrapped with calls to \verb|gettimeofday()| and setup code.
The resulting program is compiled with a C compiler for the platform under test and executed.
We compute the throughput of the experiment with the following formula:
\[
  \tpmeas{e} = \frac{\text{measured time}\times\text{frequency}}{\#\text{executed instances of }e}
\]
The reported throughput for an experiment is the median over multiple such measurements to accommodate for occasional fluctuations in the processor's clock frequency.


\subsection{Congruence Filtering}
\label{ssec:preprocessing}

In a processor microarchitecture, we expect that groups of instruction forms require the same execution resources.
Instruction forms whose operations are implemented similarly in the processor, \eg addition and subtraction, often lead to such groups.

\tool exploits these patterns to reduce the search space of the evolutionary algorithm.
It partitions the set of instruction forms into congruence classes of instruction forms that are not distinguishable with the generated experiment set.

In this partitioning, two instruction forms $i_A$ and $i_B$ are in the same class if and only if the following conditions hold:
\begin{itemize}
  \item
    $i_A$ and $i_B$ exhibit equal individual throughputs.
  \item
    Any two experiments $\{i_A \mapsto m, i_C \mapsto n\}$ and $\{i_B \mapsto m, i_C \mapsto n\}$ that combine these instruction forms with any other instruction form $i_C$ exhibit equal throughputs.
\end{itemize}
For this purpose, we consider throughputs $t_1$ and $t_2$ equal (up to measurement errors) if their symmetric relative difference is limited by a user-specified constant~$\varepsilon$, \ie if
\[
  \frac{|t_1 - t_2|}{|t_1 + t_2| / 2} < \varepsilon
\]

For each congruence class, \tool selects a representative to be included in the instruction set for the evolutionary algorithm.
The evolutionary algorithm then only needs to consider experiments that consist of these representatives.

\subsection{Evolving \PortMappings}
\label{ssec:evo}
The core of \tool is an evolutionary algorithm that searches for a \portmapping that accurately explains the observed throughputs for a given set of experiments.
Evolutionary algorithms are a well-proven technique to approach optimization problems.
They mimic concepts from natural evolution to approximatively optimize complex metrics in non-linear problem settings.
We refer to the textbook by \citet{dejong06} for a comprehensive treatment.

Every evolutionary algorithm is centered around a representation scheme that characterizes the space of possible solutions of the optimization problem.
Naturally, the scheme that we use is that of \portmappings with \uop decomposition as described in \autoref{ssec:uop_decomp}.
The sets $\Insns$ of Instructions and $\Ports$ of Ports are given by the user.
We identify each \uop with the set of ports that can execute it and allow all non-empty subsets of $\Ports$ as \uops.
The width~$|u| = |\{k \mid (u, k) \in M \}|$ of a \uop~$u$ is the number of ports that can execute~$u$.

\begin{algorithm}
  initialize population randomly\\
  \While{not done}{
    apply evolutionary operators\\
    evaluate fitness\\
    select new population\\
  }
  perform local search\\
  \Return fittest individual
  \caption{Structure of the evolutionary algorithm}
  \label{algo:evo_algo}
\end{algorithm}

\tool's evolutionary algorithm follows the structure in \autoref{algo:evo_algo}.
Initially, a set of $p$ \portmappings is sampled randomly to form a population.
This population is iteratively refined through evolution steps.
In each such step, $p$ child mappings are generated via evolutionary operators.
The resulting population of $2 p$ \portmappings is sorted according to the fitness metric and the best-performing $p$ mappings are selected as the new population.
The evolution terminates once the fitness of the population has converged to a single value or an iteration limit is exceeded.
By selecting a value for $p$, the user can find a trade-off between inference time and quality of the inferred \portmapping.

After the evolution terminates, \tool employs a greedy hill-climbing algorithm to move from the found solutions to a local optimum in the space of possible \portmappings.
It incrementally adjusts the number $n$ of \uop occurrences for each edge $(i, n, u)\in N$ and keeps the changes to the \portmapping if it is fitter than before.

In the following, we describe the components that constitute the evolutionary algorithm in detail.

\subsubsection*{Initialization}
Each member of the initial population is sampled randomly from the set of possible \portmappings as follows.
For each instruction~$i$, a random set of 1 to $|\Ports|$ many different \uops is sampled.
The number of occurrences for each of these \uops~$u$ in the mapping for~$i$ is sampled from the interval $[1, \left\lceil \tpmeas{i} \cdot |u|\right\rceil]$.
The upper bound of this interval is an implication of the throughput model:
An instruction with $\left\lceil t \cdot |u|\right\rceil$ instances of a \uop $u$ in its decomposition can achieve no throughput smaller than $t$.


\subsubsection*{Evolutionary Operators}
Evolutionary operators create new individuals from existing individuals in the population.
The most common operators in evolutionary algorithms are recombination and mutation.

We employ a binary recombination operator that mixes the information of two parent mappings to generate two child mappings.
For each instruction $i$, the set of occurring \uops with multiplicities is divided randomly into two parts that form the corresponding assignments for the children.
This operator is applied to individuals that are sampled uniformly at random from the population.

When designing the evolutionary algorithm, we tried various random mutation strategies.
Experiments showed little to no benefit over a design without a mutation operator while contributing substantial numbers of fitness computations.
Therefore, we eliminated mutation operators from our design to explore larger populations more effectively in the same execution time.

\subsubsection*{Fitness Metric}
\label{sssec:fitness_metric}
\tool's evolutionary algorithm approximately solves a multiobjective optimization problem (MOP) with the goal of minimizing two metrics:
The average relative prediction error~$D_{avg}$ and the \uop volume~$V$.
These metrics describe the quality of a \portmapping $m = (\Insns\disjointunion\Uops\disjointunion\Ports, N\disjointunion M)$ for a set $E \subseteq (\Insns \rightarrow \nat) \times \reals$ of experiments with measured throughputs as follows:
\begin{align*}
  D_{avg}(m) &= \frac{1}{|E|} \sum_{(e, t) \in E} \frac{| \tpsim{m}{e} - t |}{t}\\
  V(m) &= \sum_{(i, n, u)\in N} n \cdot |u|
\end{align*}

A low value for $D_{avg}(m)$ ensures an accurate prediction whereas a smaller \uop volume indicates a more compact and therefore more interpretable mapping.

We solve the MOP through a priori scalarization, as described \eg in Chapter 4.1 of the textbook by \citet{miettinen99}:
We combine the objectives into a single one that is interpreted as the fitness function~$F(m)$ as follows:
\[
  F(m) = \Lambda_1(D_{avg}(m)) + \Lambda_2(V(m))
\]
$\Lambda_1$ and $\Lambda_2$ are affine transformations that are chosen in every iteration to normalize both objective metrics to the range $[0, 1000]$.
They ensure that the extremal objective values of the current population are mapped to 0 and 1000, respectively, with all other objective values in between.

Combining the accuracy metric $D_{avg}$ with a compactness metric is necessary because throughput measurements usually do not uniquely identify a single \portmapping.
The \portmapping model is flexible enough to allow for a wide range of well-performing mappings with different characteristics.
While the found compact mappings are not necessarily identical to the \portmappings that are really used in the processor, they still capture the performance characteristics of the hardware as they are observable from the outside.

\subsection{Efficient Bottleneck Simulation Algorithm}
\label{ssec:ga_simulation}

Practical applicability of evolutionary algorithms depends on evaluating the fitness of many candidates in as little time as possible.
For a given time budget, fitness evaluation speed directly corresponds to the quality of the obtained solution.
With faster fitness evaluation, more candidates for survival can be considered, resulting in superior solutions.

Therefore, a critical component of our approach is the efficient simulation of experiments under a given \portmapping.
Instead of directly solving the linear program from \autoref{ssec:two_level_mapping}, we use a \bnalgo that computes the optimal solution of the linear program.
We restrict our presentation here to \portmappings in the two-level model for a more concise description.
As we have observed in \autoref{ssec:uop_decomp}, this extends to the three-level model straightforwardly.

The \bnalgo implements the following characterization of the throughput~$\tpsim{m}{e}$ of an experiment~$e$ under the \portmapping~$m := (\Insns \disjointunion \Ports, M)$:
\begin{equation}
  t_m^*(e) = \max_{Q\subseteq \Ports} \frac{\sum \{\insnf{e}(i) \mid \mathit{Ports}(m, i) \subseteq Q \}}{|Q|}
  \label{eq:bnsim}
\end{equation}
$\mathit{Ports}(m, i) := \{k \mid (i, k) \in M\}$ denotes the set of ports that can execute an instruction~$i$ under~$m$.
This characterization is based on the observation that the throughput $\tpsim{m}{e}$ has to be determined by a non-empty set $Q^*$ of bottleneck ports.
Each of the ports in $Q^*$ has to execute a mass of instructions that is equal to $\tpsim{m}{e}$.
In other words, $\tpsim{m}{e}$ is equal to the total mass of instructions that need to be executed on ports from $Q^*$, divided by the size of $Q^*$.
An optimal scheduler will assign instructions that do not need to be executed on ports from $Q^*$ to less utilized ports.
For each $Q$, the maximized term from \autoref{eq:bnsim} is a lower bound to~$\tpsim{m}{e}$.
Consequentially, finding a maximal term gives us precisely the throughput~$\tpsim{m}{e}$.
A formal proof for this equation is given in \autoref{sec:correctness}.

\begin{example}
For the execution in \autoref{fig:ex_throughput}, $Q^*$ is the set $\{P_1, P_2\}$.
  Trying to move mass from one of these ports to any other ports is either not possible (for \insn{mul}) or causes another port from $Q^*$ to execute more mass.
  $P_3$ on the other hand is irrelevant for the throughput of the experiment.
\end{example}
Our algorithmic implementation of this characterization computes the $\max$ operation in \autoref{eq:bnsim} by enumerating all subsets of the set of ports and evaluating the corresponding term.
The run-time of this algorithm is in $\Theta(2^{|\Ports|})$, which is substantially more expensive than the polynomial run-time of LP solving~\citep{bertsimas97} from a complexity-theoretic point of view.
Nevertheless, this algorithm is considerably faster for practical problems, as we show in \autoref{ssec:simulation_performance}.
On the one hand, this is due to the small number of execution ports available in modern systems.
Typical systems have eight (\eg Intel Skylake~\citep{intel19} and ARM A72~\citep{arm15}) or ten (\eg AMD Ryzen~\citep{amd17}) ports available.
On the other hand, thanks to the simplicity of the above algorithm, it is amenable to aggressive performance optimizations such as vectorization.

%% file: tex/eval.tex
\section{Evaluation}
\label{sec:eval}

This section evaluates three aspects of our work:
\begin{itemize}
  \item
    The appropriateness of the processor model as described in \autoref{sec:procmodel} and our mechanism for measuring throughput (\autoref{ssec:model_quality}).
  \item
    The quality of the inferred \portmappings for three microarchitectures from different manufacturers (\autoref{ssec:pred_quality}).
  \item
    The performance characteristics of the \bnalgo (\autoref{ssec:simulation_performance}).
\end{itemize}

\subsection{Setup}

\subsubsection{Evaluated Processors}
We use three devices with processors of distinct manufacturers for our evaluation, denoted as SKL, ZEN, and A72 in the following.
Relevant parameters are listed in \autoref{tab:procs}.
SKL has a separate pipeline of long-running operations, marked as DIV, that has to be modeled as an additional port.
One port of A72 is only used for processing branch instructions (BR). It is omitted in our model as we do not consider instructions that alter control flow.
All evaluated systems have frequency scaling and flexible overclocking mechanisms (\eg Intel Turbo Boost) disabled to facilitate reliable measurements.

\begin{table}
  \caption{Evaluated processors}
  \begin{center}
  \begin{tabular}{r|ccc}
     & \textbf{SKL} & \textbf{ZEN} & \textbf{A72}\\
    \hline
    Manufact. & Intel & AMD & RockChip\\
    Processor & Core i7 6700 & Ryzen 5 2600X \hspace{-8pt} & RK3399\\
    Microarch. & Skylake & Zen+ & Cortex-A72\\
    \# Ports & 8 + DIV & 10 & 7 + BR\\
    Instr. Set       & x86-64  & x86-64 & ARMv8-A\\
    Clock Freq. & 3.4 GHz  & 3.6 GHz     & 1.8 GHz \\
    RAM       & 32 GB   & 32GB        & 4GB\\
  \end{tabular}
  \end{center}
  \label{tab:procs}

\end{table}

A72 and ZEN are of particular interest since they do not provide the per-port performance counters that other approaches rely on \citep{amd19,arm16} whereas SKL gives means for a comparison to related work.

\subsubsection{Considered Instructions}

We select for each instruction set architecture (ISA) under test a relevant set of instruction forms.
These sets are derived from the instructions that compilers emit when compiling the SPEC CPU 2017 benchmarks\citep{spec17}.
Our instruction forms for the ARMv8-A ISA are extracted from the instructions that GCC (version 4.9.4, flags: \verb|-O3|) emits.
For x86-64, we only extract the used instruction mnemonics from the output of Clang (version 8.0, flags: \verb|-O3 -mavx2|) and use the machine-readable inputs of \citet{abel19} to generate the corresponding instruction forms.

We exclude the following instructions from these sets:
\begin{itemize}
  \item Branch and jump instructions, since their throughput heavily depends on the branch predictor.
  \item Instructions with implicitly read operands, since these cause dependencies that cannot be resolved through register allocation.
    Throughput for these could be measured by introducing additional dependency-breaking instructions as done by \citet{abel19}.
  \item x86 SSE instructions, since these add transition penalties when benchmarked together with AVX instructions.
  \item All instruction variants that operate on subregisters, to keep the run time of the evaluation bearable.
  \item x86 instructions that are not supported by Ithemal \cite{mendis19}, to have a common baseline for all comparisons.
\end{itemize}
The resulting instruction descriptions contain 310~x86-64 instruction forms and 390~ARMv8-A instruction forms.

\subsection{Processor Model and Measurements}
\label{ssec:model_quality}
In this section, we validate the practicality of the throughput model and the measurement mechanism.
We compare measured throughputs with the results of a simulation according to the processor model with a ground truth \portmapping from the work by \citet{abel19}.
Since this work only provides port usage for Intel architectures, we compare their Intel Skylake \portmapping to our measurements on SKL.
When performing this evaluation, we discovered two bugs in their \portmapping that were acknowledged by the authors.
We fixed these in the \portmapping that is used for our evaluation.

\begin{figure}[ht]
  \includegraphics{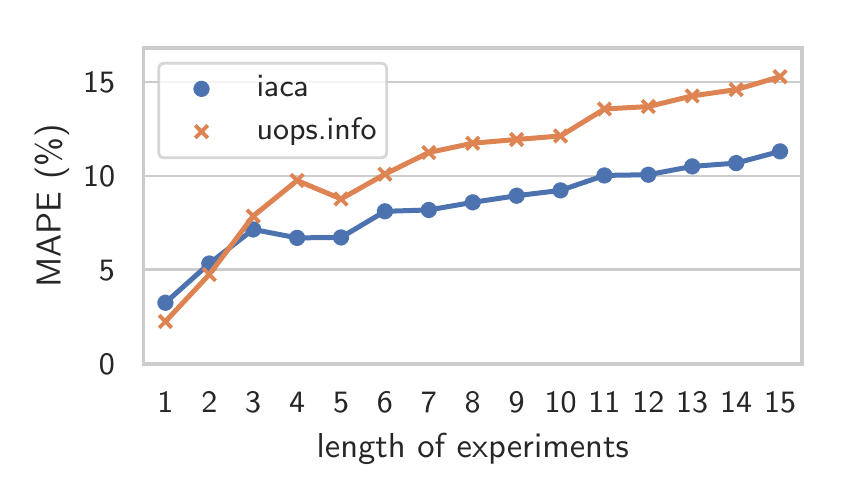}

  \vspace{-10pt}
  \Description{
    Mean absolute percentage error (MAPE) of simulation with the \portmapping from \citet{abel19} and with IACA \citep{iaca} with respect to our measurements for experiments of varying length.
    Both lines show similar behavior.
    Starting from experiments with length 1, both lines rise from circa $3\%$ at first up to around $8\%$ at 3 instructions.
    From this point, with a few outliers, the lines rise close to linearly, but less steep than before, ending at 15 instructions with $11\%$ for IACA and $15\%$ for uops.info.
  }
  \caption{Mean absolute percentage error (MAPE) of simulation with the \portmapping from \citet{abel19} and with IACA \citep{iaca} with respect to our measurements for experiments of varying length}
  \label{fig:varying_length}
\end{figure}

\autoref{fig:varying_length} shows the mean absolute percentage error for the simulation with the \portmapping from Abel and Reineke with respect to our measurements for varying experiment length.
For length 1, we use the set of all supported x86-64 instructions, whereas for larger lengths, we randomly sample 2,000 experiments from the set of all instruction multi-sets of the appropriate size.

For small experiment lengths, we can see a low error showing that the experiments behave as predicted by the processor model.
With increasing length of experiments, the accuracy degrades.
The lower prediction error of IACA \citep{iaca} in \autoref{fig:varying_length} indicates that with longer experiments, the influence of factors such as non-optimal scheduling decisions that are not covered in the throughput model (but by IACA) rises.

Overall, the error is small enough to justify the use of measurement mechanism and throughput model.

\subsection{Model Predictions}
\label{ssec:pred_quality}

Directly measuring the quality of a \portmapping is hindered by the lack of ground truth for most processors.
We therefore assess the inferred \portmappings by their ability to accurately predict the measured throughput of \portmappingbound experiments.
For each microarchitecture, we use a different benchmark set of 40,000 experiments, which we instantiate with operands and whose throughput we measure as described in \autoref{ssec:expeval}.
These experiments are sampled uniformly at random from the set of all instruction multi-sets of size 5.

One major use case of \tool is to provide port mappings for performance estimation tools.
Therefore, we compare the prediction accuracy of \tool's mappings to the modeling of \portmappings in state-of-the-art performance prediction tools.
To this end, we use the same benchmark sets to evaluate IACA \citep{iaca} (version 3.0), llvm-mca \citep{llvmmca} (from LLVM version 8.0.1\footnote{Initially, we performed these experiments on the more recent LLVM version 9.0.1 but found a severe regression in prediction accuracy on our experiments compared to version 8.0.1.}), Ithemal \citep{mendis19}, and the \portmapping provided by uops.info \citep{abel19} for their respective supported platforms.
Note that these benchmarks specifically stress the port-mapping aspect of these prediction tools because they do not contain any data dependencies.
They are therefore not representative to evaluate the overall prediction quality of these tools on compiler-generated code.\footnote{We refer to the BHive project \citep{chen19} for an evaluation of their accuracy for instruction sequences extracted from code generated for common benchmarks.}
\autoref{sec:relwork} discusses the performance estimation tools we evaluate in further detail.

Of these four related approaches, only the \portmapping from uops.info is directly comparable to \tool{}'s results because it can only predict the throughput of instruction sequences without data dependencies.
The other approaches are more general in that they can predict the throughput of arbitrary instruction sequences, but might not be attempting to provide good accuracy for dependency-free code.
For example, Ithemal uses a neural network model trained via supervised learning rather than an explicit \portmapping model.
Being trained on collected basic blocks from entire programs where dependencies are to be expected, accurate predictions for dependency-free code might be outside of the scope of Ithemal.

For all three platforms, we ran our \tool prototype with a population size of 100,000 and an $\varepsilon$ of $0.05$ for congruence filtering.
\autoref{tab:characteristics} gives numbers on the time required to benchmark throughputs for experiments and to infer a \portmapping for all considered platforms.
It further shows that the effectiveness of congruence filtering is considerable: The relevant instructions are reduced by $53\%$ to $69\%$.
The low number of different \uops used in the inferred \portmappings indicates that \tool developed compact representations for all three platforms.
The uops.info \portmapping for SKL uses 12 different \uops for the same set of instructions.

\begin{table}[ht]
  \caption{\tool mapping characteristics}
  \begin{center}
  \begin{tabular}{r|ccc}
    & \textbf{SKL} & \textbf{ZEN} & \textbf{A72}\\
    \hline
    benchmarking time & 20h & 27h & 74h \\
    inference time & 5h & 21h & 12h \\
    insns found congruent & $69\%$ & $53\%$ & $56\%$ \\
    number of \uops & 17 & 15 & 9 \\
  \end{tabular}
  \end{center}
  \label{tab:characteristics}
\end{table}

To provide a broad comparison of prediction accuracy, we give results for the following commonly used accuracy metrics:
\begin{itemize}
    \item
  The Mean Absolute Percentage Error (MAPE) is a measure of the relative error of the simulation over measurements.

\item
  The Pearson Correlation Coefficient (PCC) describes how closely the relation between simulation and measurements can be described by a linear equation.

\item
  The Spearman Correlation Coefficient (SCC) is a measure of rank correlation.
  A high rank correlation indicates that if the measurement for one experiment is smaller than for another experiment, its simulated value is likely to be smaller as well.
\end{itemize}
The value range for PCC and SCC is $[-1,1]$, ranging from negative correlation ($-1$) over no correlation ($0$) to maximal correlation ($1$).

\begin{figure*}[ht]
  \includegraphics{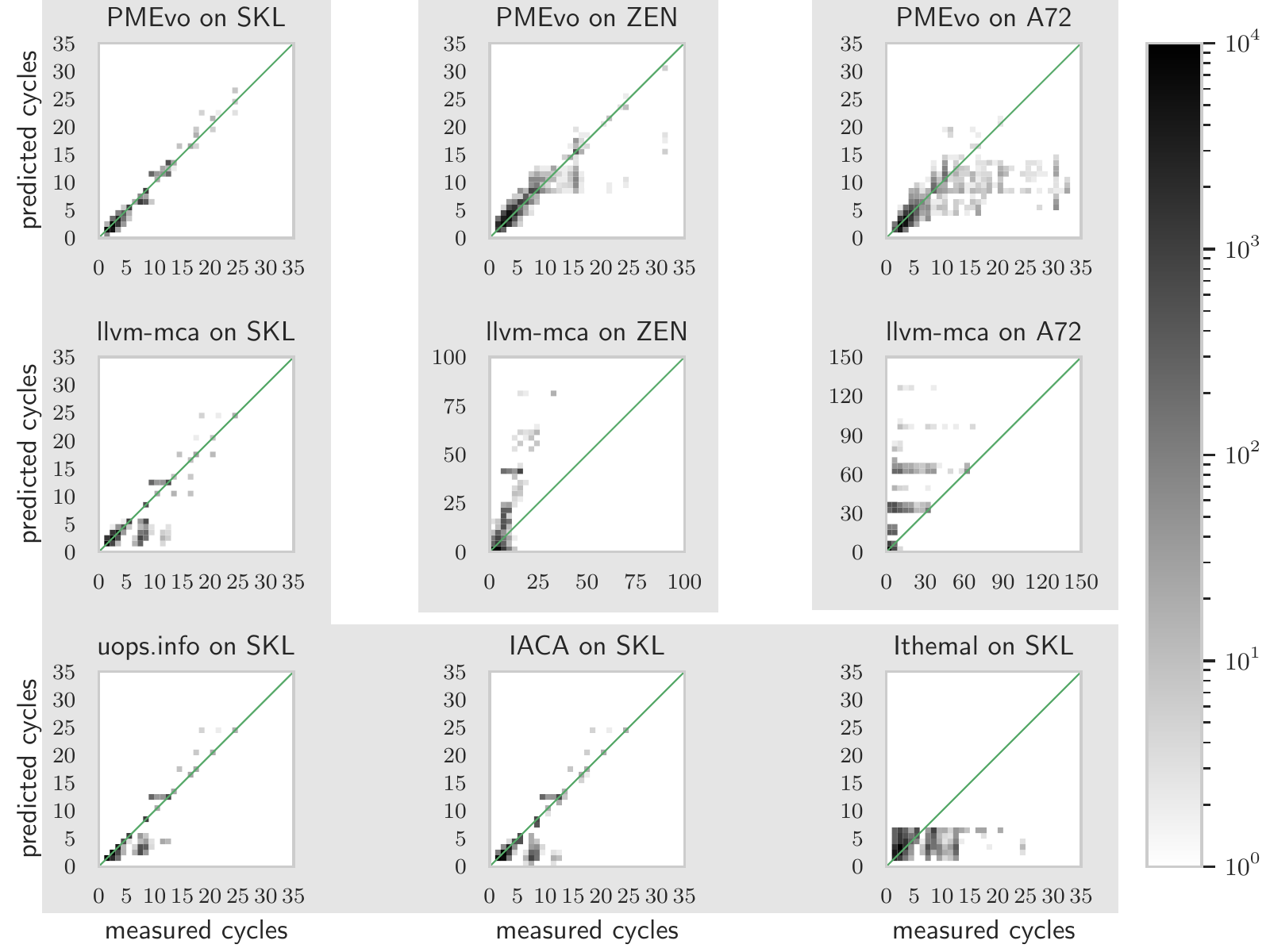}
  \Description{
    Prediction accuracy on port-mapping-bound experiments.
    Each heat map relates predicted and measured throughput in cycles per experiment.
    Points closer to the diagonal line indicate better predictions.
    Please refer to the section text for a detailed discussion of this figure.
  }
  \caption{
    Prediction accuracy on \emph{\portmappingbound experiments}.
    Each heat map relates predicted and measured throughput in cycles per experiment.
    Points closer to the diagonal line indicate better predictions.
    The gray boxes group heat maps for the same platform.
    The experiments were set up and measured as described in \autoref{ssec:expeval}.}
  \label{fig:accuracies}
\end{figure*}

Additionally, we visualize the prediction accuracy of our approach in comparison to related work in \autoref{fig:accuracies} with a heat map for each pair of architecture and prediction mechanism.
For each heat map, the experiments are considered as data points with measured and predicted throughput.
The heat map shows the space of possible pairs of measured and predicted throughput, split into $35\times35$ equally sized bins.
Each bin's shade represents the number of experiments that lie in it.
Ideally, measurement and prediction agree, leading to experiments close to the marked diagonal line.
Experiments below the diagonal indicate an under-estimation of the throughput, those above are over-estimated by the predictor.

We discuss the represented data in detail in the following sections.

\subsubsection{SKL}
For the Intel Skylake platform, we compare the prediction accuracy of \tool to all afformentioned approaches: the \portmapping from uops.info, IACA, llvm-mca, and Ithemal using its publicly-available pre-trained network for the Skylake microarchitecture.

The inputs for IACA, llvm-mca, and Ithemal consist of the loop body of the experiments, unrolled to a length of ten instructions so that operand allocation can avoid loop-carried dependencies.
For the entire set of experiments, we report the results of the tools for this input, divided by the number of experiments in the unrolled loop body.

The accuracy metrics for the five tools under comparison are listed in \autoref{tab:accuracy}.

\begin{table}[ht]
  \caption{Prediction accuracy measures for \portmappingbound experiments on SKL}

  \vspace{-4pt}
  \begin{center}
  \begin{tabular}{r|ccc}
    & MAPE & Pearson CC & Spearman CC\\
    \hline
    \tool & $14.7\%$ & $0.98$ & $0.85$ \\
    uops.info & $9.3\%$ & $0.92$ & $0.88$ \\
    IACA & $8.0\%$ & $0.86$ & $0.79$ \\
    llvm-mca & $9.7\%$ & $0.87$ & $0.82$ \\
    Ithemal & $60.6\% $ & $0.35$ & $0.54$ \\
  \end{tabular}
  \end{center}
  \label{tab:accuracy}

  \vspace{-6pt}

\end{table}

IACA, llvm-mca, and uops.info all predict with an average error of less than $10\%$ with high correlation values.
This impression is confirmed by the corresponding heat maps in \autoref{fig:accuracies}: Most of the experiments are close to the ideal line.
They also all show a cluster of  experiments below the diagonal line.
These can be attributed to the family of bit test instructions (BTx), for which the measurable throughput does not agree with the throughput implied by the port usage as confirmed by the measurements of \citet{abel19}.

Our approach, \tool, has a slightly higher relative error than IACA, llvm-mca, and and uops.info, but comparable correlation coefficients.
The corresponding heat map in \autoref{fig:accuracies} shows a distribution close to the diagonal line.
The BTx instructions that caused inaccuracies for the other approaches have a representation as multiple \uops that map to the same ports.
While differing from the real \portmapping, this fits better to the observable throughputs.

\begin{figure*}[ht]
  \centering
  \begin{subfigure}{.47\textwidth}
    \includegraphics{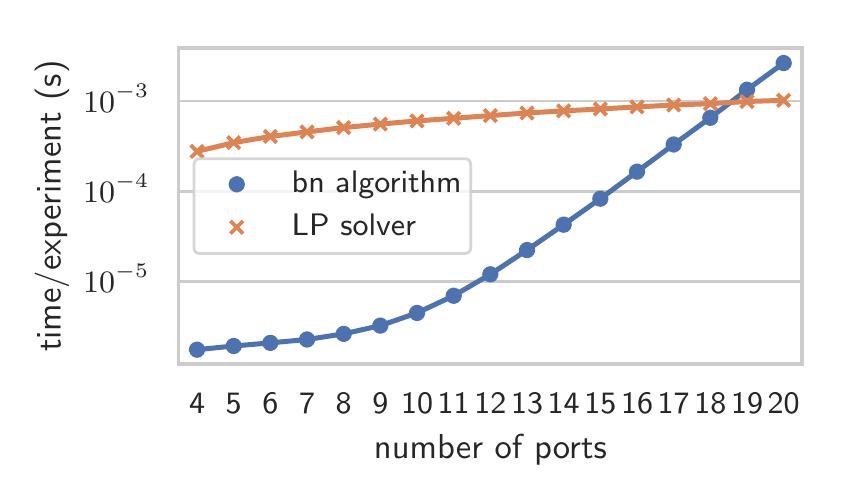}

    \vspace{-10pt}
    \subcaption{}
    \label{fig:simulator_comparison_ports}

    \vspace{-10pt}
  \end{subfigure}
  \begin{subfigure}{.45\textwidth}
    \includegraphics{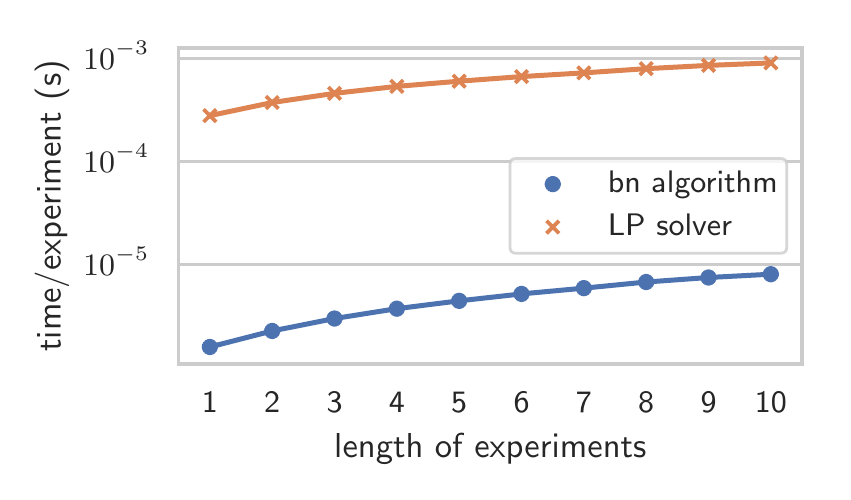}

    \vspace{-10pt}
    \subcaption{}
    \label{fig:simulator_comparison_len}

    \vspace{-10pt}
  \end{subfigure}
  \Description{
  Execution time comparison of the \bnalgo and the LP solver with varying port numbers with experiments of length 4 (\subref{fig:simulator_comparison_ports}) and with varying length of experiments with 10 ports (\subref{fig:simulator_comparison_len}). Both have their vertical axis in a logarithmic scale.
  In both images the LP solver curve is rising sublinearly (in log scale), from above $10^{-4}$ seconds to $10^{-3}$ seconds.
  Also in both images, the bottleneck algorithm's curve starts at $10^{-6}$ seconds and starts by rising similarly slow as the LP solver curve.
  For the length of experiments, this trend continues through the entire figure up to 10 instructions.
  The bottleneck algorithm's curve for the number of ports becomes steeper between 9 and 12 ports, from where it continues linearly (in log scale).
  It intersects with the LP solver curve between 18 and 19 ports.
  }
  \caption{Execution time comparison of the \bnalgo and the LP solver with varying port numbers with experiments of length 4 (\subref{fig:simulator_comparison_ports}) and with varying length of experiments with 10 ports (\subref{fig:simulator_comparison_len}). Both have their vertical axis in a logarithmic scale.}
  \label{fig:simulator_comparison}
\end{figure*}

For Ithemal, we observe lower correlations and a high error rate.
This differs from the evaluation by \citet{mendis19} where Ithemal exhibits superior results in these metrics in comparison to IACA.\footnote{Their findings for the accuracy of IACA are consistent with the ones presented here.}
As already noted, the difference in performance is likely a consequence of the different characteristics of the experiments used here and in the experimental evaluation of their paper:
Ithemal is trained and validated on basic blocks emitted from a compiler for entire programs, which exhibit substantially more data dependencies than our experiments.

However, an appropriate interpretation of these results needs to be judicious:
A high prediction accuracy for our experiments could have indicated a generalization of Ithemal to dependency-free code.
Yet, the observed low prediction accuracy for our inputs does not allow conclusions about Ithemal's performance across real-world programs.

\subsubsection{ZEN and A72}
For the AMD and ARM microarchitectures, we compare \tool's results only to llvm-mca since the other approaches are only available for Intel architectures.

The metrics for both architectures in \autoref{tab:accuracy_other} show a common trend:
\tool exhibits a considerably smaller prediction error than llvm-mca.

\begin{table}[t]
  \caption{Prediction accuracy measures for \portmappingbound experiments on ZEN and A72}
  \begin{center}
  \begin{tabular}{r|ccc}
    & MAPE & Pearson CC & Spearman CC\\
    \hline
    \tool (ZEN) & $13.5\%$ & $0.94$ & $0.87$ \\
    llvm-mca (ZEN) & $50.8\%$ & $0.86$ & $0.54$ \\
    \hline
    \tool (A72) & $21.4\%$ & $0.68$ & $0.77$ \\
    llvm-mca (A72) & $65.3\%$ & $0.67$ & $0.68$ \\
  \end{tabular}
  \end{center}
  \label{tab:accuracy_other}
\end{table}

For ZEN, \tool inferred a \portmapping that predicts with close to equal accuracy as its SKL mapping.
With $21.4\%$, the prediction error of the \tool mapping for A72 is notably higher while correlations are lower.
This observation is confirmed by the corresponding heat maps in \autoref{fig:accuracies}.
\tool on A72 is prone to under-estimating experiments with longer running times.
We attribute this to A72's less advanced out-of-order execution engine (according to the respective optimization guides \citep{arm15,amd17,intel19}), which renders the experiments less representative for the \portmapping.

In contrast to its results for SKL, llvm-mca has substantially larger prediction errors.
The heat maps indicate a significant over-estimation of the throughput.
One possible explanation is that these architectures are less in the focus of the developers than SKL and the respective \portmapping models might not yet be as elaborate and accurate as the one for SKL.
Especially for these two architectures, the models derived with \tool may significantly increase the accuracy of llvm-mca's throughput prediction.

\subsection{Performance of the Simulation Algorithm}
\label{ssec:simulation_performance}

This section explores the performance behavior of the \bnalgo as presented in \autoref{ssec:ga_simulation}.
For this purpose, we compare our optimized implementation of the \bnalgo to a realization of the linear program from \autoref{ssec:uop_decomp} in the state-of-the-art LP solver Gurobi~\citep{gurobi19} (version 7.5.2).
The running times reported for the LP version include model construction via the Gurobi C++ API as well as the actual solving.

There are two significant parameters that influence the execution time of both simulation methods: the number of ports in the microarchitecture and the length of the experiments.\footnote{The number of instructions in the instruction set architecture is not relevant, since both implementations ignore instructions that do not occur in the experiment.}
We evaluate these parameters with randomly generated microarchitectures with the appropriate number of ports for an artificial instruction set architecture of 100 instructions.
For each (number of ports, length of experiments) configuration, 128 randomly sampled experiments were simulated with each of 8 randomly sampled three-level mappings.
The resulting seconds per experiment value for each pair of experiment and mapping is the arithmetic mean over 1000 simulations.
The points in the graph mark the median of these values for each (number of ports, length of experiments) configuration.

\paragraph{Influence of the Number of Ports}
\autoref{fig:simulator_comparison_ports} shows the results for experiments consisting of 4 instructions with a varying number of ports.
For port numbers up to 10 as they occur in contemporary platforms, the \bnalgo outperforms the linear program by two orders of magnitude.
Starting from 12 ports, the simulation time with the \bnalgo rises with a stronger incline.
The \bnalgo reaches the simulation time of the LP implementation at about 18 ports.
With the same inputs, the simulation time via the LP solver grows substantially slower with the number of ports.
We conclude that the exponential run-time behavior of the \bnalgo, as explained in \autoref{ssec:ga_simulation}, has a negligible impact for inputs of interest.




\paragraph{Influence of the Length of Experiments}
The experiments we use for the evolutionary algorithm are of very limited length to allow reliable execution on actual processors.
Nevertheless, exploring the behavior with different lengths of experiments is worthwhile for the discussion of the \bnalgo.
The results for varying lengths of experiments in an architecture with 10 ports are displayed in Figure~\ref{fig:simulator_comparison_len}.
Here, the \bnalgo consistently outperforms the LP solver by two orders of magnitude.
The execution time for both methods grows sub-exponentially with the length of experiments, with an almost identical incline in the log-scale plot.
This indicates that the rate at which the execution time rises with growing experiment length for the LP solver is considerably higher than for the \bnalgo.

%% file: tex/relwork.tex
\section{Related Work}
\label{sec:relwork}

We divide related work into two categories:
Approaches to find \portmappings and work on predicting instruction throughput.

\subsection{Inferring \PortMappings from Experiments}

The instruction tables by \citet{agner18} used to be the only available source for experimentally validated information on instruction latency, throughput, and port usage.
They are obtained with hand-crafted microbenchmarks that use hardware performance counters to count the number of executed cycles and the number of executed \uops per port.
\citet{abel19} show that the reported port usage by Fog is only an under-approximation of the usable ports.

For the case that these counters are not available, Fog uses experiments that execute instructions with unknown port usage together with instructions whose port usage is known from some other resource.
Observing the running time allows to identify interfering instruction combinations.

The tables include such information for a wide range of x86 microarchitectures by Intel, AMD, and VIA.
The requirement to construct suitable microbenchmarks for each microarchitecture makes this approach very work-intensive.

\citet{abel19} automated the process of designing microbenchmarks to measure latency, throughput, and port usage.
Their algorithm to estimate port usage overcomes the imprecision of Fog's approach by using blocking instructions.
The processor decomposes these instructions each into a single \uop that can only be executed on a known set of ports.
When executing the instruction under test with a sufficient number of blocking instructions to fully saturate a set $P$ of ports, \uops of the instruction under test will be executed on ports not in $P$ if possible.
For observing this as well as for identifying blocking instructions, they use per-port hardware performance counters as they are used by \citet{agner18}.
While they provide throughput and latency measurements for x86 microarchitectures by Intel and AMD, they only give \portmappings for the Intel platforms as only these provide all required performance counters.

Two further approaches initiated by Google are collected under the name EXEgesis.
One is the EXEgesis project~\citep{exegesis} that extracts latencies, throughputs, and port usage for Intel architectures from vendor-provided documentation.
This requires automatically parsing documents that were intended for human readers: a fragile and work-intensive process.
Since the provided documentation does not include all relevant information, the EXEgesis developers also created tools to infer the missing information via experiments.
This led to the second project under this name, llvm-exegesis~\citep{llvmexegesis}, a tool inside the LLVM framework~\citep{lattner04} that automatically generates benchmarks similar to those used by~\citet{agner18}.
For measuring port usage, llvm-exegesis depends on per-port performance counters just as the two previously discussed approaches.

All of these works compare to ours in a similar way:
Since they use precise hardware performance counters, they can obtain more accurate \portmappings than our approach.
However, our approach does not suffer from the restriction to platforms that have these performance counters, allowing us to automatically infer \portmappings for x86 platforms by AMD, as well as for ARM platforms.

\subsection{Work on Instruction Throughput Prediction}
As \portmappings are commonly used for throughput prediction, it is instructive to set the presented results in context to work from this field.

The Intel Architecture Code Analyzer (IACA) \citep{iaca} models the execution of a sequence of instructions, considering factors such as port usage, operand dependencies, and instruction decoding bottlenecks.
The output of IACA for a given instruction sequence includes a throughput estimation, a bottleneck resource, and the distribution of \uops to ports.
It is a closed-source tool provided by the processor manufacturer Intel for some of its microarchitectures.
As a consequence, IACA can make use of unpublished internal information to achieve an accurate performance prediction.
Nevertheless, previous research (\eg \citep{abel19}) has shown cases where the prediction of IACA differs from the observable behavior.
Since April 2019, IACA is no longer under active development.

OSACA \citep{laukemann18} is an attempt to provide the same features as IACA, but with a non-proprietary system.
They use information from \portmappings for their supported architectures, a range of Intel microarchitectures as well as AMD's Zen architecture.
These \portmappings are extracted from sources like the tables by \citet{agner18} and material provided by the manufacturers.
They implement means of experimentally validating this known port model via experiments, noting that experiments with multiple different instructions can uncover new details of the \portmapping.
Our approach systematically extends this line of work to derive new \portmappings.

The llvm-mca tool \citep{llvmmca} is also inspired by IACA.
It uses knowledge from the LLVM \citep{lattner04} instruction scheduling models, including port usage if available, for performance prediction.
These scheduling models are the result of human fine-tuning effort, proprietary knowledge contributed by processor designers, and experiments via llvm-exegesis \citep{llvmexegesis}.

Both, llvm-mca and OSACA, can benefit from \portmappings by \tool for microarchitectures without available \portmapping.

Ithemal \citep{mendis19} uses machine learning techniques for instruction throughput prediction.
Similar to our approach, it only needs a specification of the instruction set architecture under test and a set of experiments labeled with measured throughputs as an input.
These labeled inputs are used as training data for a hierarchical recurrent neural network based on long short-term memory (LSTM) cells.

Ithemal is trained and validated on basic blocks that are extracted from compiled benchmark programs.
As a result, Ithemal captures different aspects than our approach:
\tool focuses on experiments whose outcome is solely determined by the \portmapping whereas the predictions of Ithemal are shaped by other factors such as data dependencies.

A drawback of the Ithemal approach is that the resulting processor model can only be interpreted by evaluating it on an instruction sequence.
This is sufficient for certain applications like stochastic superoptimization \citep{schkufza13}.
However, in applications like the backend of an optimizing compiler, enumerating and evaluating a large set of possible instruction sequences is prohibitively expensive.
A compact \portmapping is more easily interpreted for constructing well-performing instruction sequences as it clearly indicates which instructions have conflicting resource requirements.

%% file: tex/conclusion.tex
\section{Conclusion}
\label{sec:conclusion}

This paper presents \tool, a framework to infer \portmappings, \ie compact and interpretable representations of a modern processor's ability to exploit instruction-level parallelism.
The inference is done by an evolutionary algorithm that optimizes \portmappings to explain the instruction throughputs measured for specifically designed instruction sequences.
Using a novel \bnalgo to evaluate the fitness of \portmappings, \tool can explore the large search space of possible mappings effectively.

We demonstrate \tool's portability by inferring \portmappings for three different microarchitectures, two of which are out of scope for previous automatic approaches.
The high prediction accuracy of the inferred \portmappings shows that \tool can make performance engineering tools more reliable for a wide range of hardware platforms.

%% file: tex/correctness.tex
\section{Correctness of the Bottleneck Simulation Algorithm}
\label{sec:correctness}

The proof of correctness of the bottleneck simulation algorithm presented here uses basic results from linear optimization theory.
For background and proofs on these results, we refer to the textbook by \citet{bertsimas97}.

Let $S(m, e)$ be defined as follows:
\[
  S(m, e) := \Bigg\{\frac{\sum \{\insnf{e}(i) \mid \mathit{Ports}(m, i) \subseteq Q \}}{|Q|}~\Big|~Q\subseteq P\Bigg\}
\]
With this notation, \autoref{eq:bnsim} can be written as $\hat{t}_m(e) = \max S(m, e)$.
The proof proceeds by showing that the result $\hat{t}_m(e)$ of the \bnalgo is equal to the throughput $\tpsim{m}{e}$ according to \autoref{def:tp2l} for any experiment $e$ and any (two-level) \portmapping $m := (\Insns \disjointunion \Ports, M)$.
We do so by showing that (I) $\tpsim{m}{e}$ is included in $S(m, e)$ and that (II) each element of $S(m, e)$ is upper-bounded by $\tpsim{m}{e}$.

\paragraph{I}
Let $s$ be an optimal feasible solution of the linear program.
We denote the value of a variable $x$ chosen in $s$ by $s[x]$.
Since $s$ is optimal, there is a non-empty maximal set $Q\subseteq P$ such that for all $k\in Q$ holds that
\begin{equation}
  \label{eq:defq}
  \sum_{i\in\Insns}s[x_{ik}] = s[t] = \tpsim{m}{e}
\end{equation}
Without loss of generality, we assume that each instruction that $s$ executes on a port in $Q$ can only be executed on ports in $Q$, that is:
\begin{equation}
  \label{eq:wlog}
  k\in Q \land s[x_{ik}] > 0 \Rightarrow \mathit{Ports}(m, i) \subseteq Q
\end{equation}
If this is not the case for $s$, we can find a different solution $s'$ with identical objective value that fulfills this constraint as follows:
For every $(i, k)$ such that $s[x_{ik}] > 0$, $\sum_{i\in\Insns}s[x_{ik}]=s[t]$, and $Q' := \mathit{Ports}(m, i) \cap (P\backslash Q) \neq \emptyset$, we remove a sufficiently small part of the value for $x_{ik}$ and add it to the value of some $x_{ik'}$ with $k' \in Q'$ such that constraint (\ref{jsB}) is tight for neither of $k$ and $k'$.\footnote{If this was possible for all $k\in Q$, $s$ could not be optimal. }

By defining $J := \{i \mid \mathit{Ports}(m, i) \subseteq Q\}$, we identify the following equalities:
\begin{align*}
  \sum_{i\in J} e(i) &\stackrel{(\ref{jsA})}{=} \sum_{i\in J} \sum_{k\in\Ports} s[x_{ik}]
  \stackrel{(\ref{jsD})}{=} \sum_{i\in J} \sum_{k\in Q} s[x_{ik}]
  = \sum_{k\in Q} \sum_{i\in J} s[x_{ik}]\\
  &\stackrel{(\ref{eq:wlog})}{=} \sum_{k\in Q} \sum_{i\in \Insns} s[x_{ik}]
  \stackrel{(\ref{eq:defq})}{=} \sum_{k\in Q} s[t]
  = \tpsim{m}{e} \cdot |Q|
\end{align*}
The equality of the leftmost term and the rightmost term proves that $\tpsim{m}{e} \in S(m, e)$.

\paragraph{II}
Let $Q'\subseteq\Ports$ and $t' := \sum \{\insnf{e}(i) \mid \mathit{Ports}(m, i) \subseteq Q \} / |Q|$.
We assume $t' > \tpsim{m}{e}$ and show that this leads to a contradiction, proving that $\tpsim{m}{e}$ is an upper bound to each element of $S(m, e)$.

For this argument, we form the dual of the linear program:
\begin{alignat*}{3}
  &\text{maximize}& \quad & \sum_{i\in\Insns} e(i) \cdot y_i \\
  &\text{subject to}& \quad & y_i - z_k \leq \overline{m}_{ik} && \fspace\text{for all $i\in\Insns$, $k\in\Ports$}\\
  &&& \displaystyle\sum_{k\in\Ports} z_{k} = 1 &&\\
  &&& z_{k} \geq 0 && \fspace\text{for all $k\in\Ports$}\\
  &&& y_{i} \geq 0 && \fspace\text{for all $i\in\Insns$}
\end{alignat*}
Here, the $y_i$ and $z_k$ are real-valued variables and $\overline{m}_{ik} = 1 \Leftrightarrow (i, k) \not\in M$.

By the strong duality theorem for linear programs, an optimal solution for this dual linear program has the same objective~$\tpsim{m}{e}$ as an optimal solution for the primal linear program.

Given the assumption that $t' > \tpsim{m}{e}$, we construct a solution~$s'$ for the dual with a higher objective value, which contradicts the strong duality theorem or the optimality of $\tpsim{m}{e}$.
The construction of $s'$ is as follows for each $i\in \Insns$ and $k\in\Ports$:
\begin{align*}
  s'[z_k] &= 1/|Q'| & \text{if $k\in Q'$}\\
  s'[y_i] &= 1/|Q'| & \text{if $\mathit{Ports}(m, i)\subseteq Q'$}
\end{align*}
All other variables are set to 0.
This solution fulfills all constraints and has the following objective value:
\[
  \sum_{i\in\Insns} e(i) \cdot y_i = \frac{\sum \{\insnf{e}(i) \mid \mathit{Ports}(m, i) \subseteq Q \}}{|Q|} = t' > \tpsim{m}{e}
\]

\vspace{10pt}
\noindent
This proves the correctness of the bottleneck simulation algorithm.\hfill$\square$